\documentstyle[12pt]{article}
\begin{document}
\newcommand{\eq}{\begin{equation}}
\newcommand{\en}{\end{equation}}
\hspace*{1cm}
{\Large\bf Weak Decays of Heavy Baryons in the\\[2mm]
\hspace*{1cm}
Covariant Quasipotential Approach}\\[2mm]
\hspace*{1cm}
{\it
     \underline{A.\ G.\ Rusetsky$^{a,c}$},
     M.\ A.\ Ivanov,$^a$
     J.\ G.\ K\"{o}rner,$^b$
     V.\ E.\ Lyubovitskij,$^{a,d}$
\\[2mm]}
\hspace*{1cm}
{\small $^a$JINR, Dubna, $^b$Mainz University,
$^c$Tbilisi University, $^d$Tomsk University}

\abstract

\noindent
Bethe-Salpeter approach has been applied to the study of $b\rightarrow c$
transitions both for heavy mesons and heavy baryons. Meson and baryon IW
functions are calculated on the equal footing. A reasonable agreement
with the experimental data for heavy to heavy semileptonic transitions
has been obtained.

\vspace*{.3cm}
\noindent
The remarkable progress in the Heavy Quark Effective Theory (HQET) $^1$
still leaves a large room for the application of model studies of
the momentum dependence of heavy hadron decay formfactors
which, being governed by the dynamics of light constituents, can not
be determined from the first principles in QCD and HQET.
(for the review of the model approaches see e.g. $^{1,2}$)
The nonrelativistic models, though very simple and transparent, in general
are known not to suffice for this purpose.
The Bethe-Salpeter (BS) approach provides a completely
relativistic treatment of the bound-state problem, in which the interaction
operators between the particles can be constructed from
the underlying field-theoretical Lagrangian. On the other hand, the BS approach
in the instantaneous approximation shares most of the transparency
of the nonrelativistic picture, and the corrections emerging due to the
deviation from the instantaneous limit are calculable by use of the quasipotential
method $^3$. For this reason, as well as in the view of the fact that
the instantaneous BS kernels (both for $q\bar q$ and $3q$ cases) have become
recently available from QCD-based calculations $^4$, a systematic
reexamination of the weak decay characteristics simultaneously in the
heavy meson and heavy baryon sectors seems to be very important.
Most of the recent studies on the subject, however, have been restricted
to the meson sector only (see, e.g. $^{5,6}$), where the comprehensive analysis,
based on the BS equation, is carried out both in the heavy quark limit $^5$
as well as including leading $1/m_Q$ corrections $^6$. The investigations of
heavy baryon decays mostly utilize the quark-diquark approximation $^7$, which
vastly reduces the complexity of the problem and allows for the use of the
technique developed for the description of two-body bound states within
BS approach. The quantitative treatment of the weak decays of the genuine
three-body bound systems, however, is still desirable.

Due to the extreme complexity of the above stated problem (especially,
bearing in mind the necessity of having to include the nonfactorizing
contributions in heavy baryon nonleptonic decays $^{8,9}$), one has to make some
simplifying assumptions which, presumably, may be lifted at the subsequent
stages of the investigation. Namely, we assume:

$\bullet$ Instantaneous interquark interactions in the c.m. frame of the
hadron (the retardation effect can be studied perturbatively, with the
use of the quasipotential method).

$\bullet$ Spectator approximation for quarks which implies the negligence
of the spin-spin interaction between constituents $^8$ (for the Lagrangian
formulation of the Spectator Model see Refs. $^9$). The BS approach
provides an appropriate tool for the study of spin interaction effects,
which can be embedded in the present framework, after the complicated
three-body dynamics in heavy baryon decays is well understood.

$\bullet$ "Mass shell" approximation for quarks which implies the fixing
of the momenta of the constituent quarks in the heavy hadron w.f.-s
on the mass shell
and seems to be natural in the spectator picture. This approximation allows
one to arrive at a very simple expression for various decay matrix elements
which are given in the form of the quantum-mechanical overlap integrals
containing equal-time (ET) w.f.-s of hadrons.

In the construction of matrix elements we abandon the "two-tier" approach $^{10}$
which is schematically given as:
\begin{center}
3D Equation $\rightarrow$ Equal-time w.f. $\rightarrow$ BS w.f. $\rightarrow$ Amplitudes
\end{center}
Though very successful and widely used in the construction of the two-body
bound-state transition amplitudes $^{5,6}$, the above approach, in our opinion,
reveals some undesirable properties when applied to the three-body bound
states. Namely, the BS w.f. constructed within this approach contains the
ill-defined square root of the $\delta$-function $^{10}$. For this reason we
prefer to work entirely in terms of the ET w.f. within the covariant
quasipotential approach.

We choose the instantaneity anzats for the three-particle pairwise
kernel ac\-cor\-ding to Ref. $^{10}$. The different choice $^{11}$, in our opinion,
seems to be less natural in the view of the underlying field-theoretical
content. In the spectator approximation both forms, however, lead to
the similar results.

At the present stage we apply the covariant quasipotential approach to the
study of $b\rightarrow c$ semileptonic transitions in the heavy quark limit.
Under the approximations listed above the expression for the heavy baryon
IW function can be easily reduced to the form (details can be found in our
forthcoming publication)
\eq
\xi_B(\omega)=\frac{1+\omega}{2}\int\prod_{i=1}^2\frac{d^3\vec p_i}{(2\pi)^3}\,\,
\phi_B(\{\vec p_i\})\,\phi_B(\{\vec p_i^{\,\prime}\})
\en
\noindent
where $\vec p_i^{\,\prime}=\vec p_i+\vec v\sqrt{m_i^2+\vec p_i^{~2}}
+\vec v(\vec v\cdot\vec p_i)/(\omega+1)$, $\,\,\omega=\sqrt{1+\vec v^{~2}}$
and $i=1,2$ corresponds to the light constituents of the heavy hadron.
The baryon ET w.f. $\phi_B$ is normalized to unity.

If one assumes that two light constituents in the heavy baryon do not
interact, the simple result is obtained
\eq
\xi_B(\omega)=\frac{1+\omega}{2}\,\,\xi_M^{(1)}(\omega)\,\,\xi_M^{(2)}(\omega)
\en
\noindent
where $\xi_M^{(i)}(\omega)$ denote the meson IW functions corresponding to
the light constituents $1$ and $2$, respectively. The expression for
it is similar to (1)
\eq
\xi_M^{(i)}(\omega)=\int\frac{d^3\vec p}{(2\pi)^3}\,\,
\phi_M^{(i)}(\vec p)\,\,\phi_M^{(i)}(\vec p^{\,\prime})
\en
Below we present the results of the fit to the baryon and meson observables
within the BS approach (spectator picture). For simplicity, at the first
step we have used the oscillator w.f.-s $\phi\sim\exp(-\vec p^{~2}/\Lambda^2)$
with the same cutoff parameter $\Lambda=500~MeV$ in the meson and baryon
sectors. The light constituent quark mass was taken to be $m=250~MeV$
(in the rough estimate we use the same masses for $u,d,s$ quarks).
The meson IW function calculated in our approach is
well approximated by
\eq
\xi_M(\omega)=\biggl(\frac{2}{1+\omega}\biggr)^{2\rho_M^2}\quad\quad
{\mbox{with}}\quad \rho_M^2=\frac{3}{4}+\biggl(\frac{m}{\Lambda}\biggr)^2
\en
\noindent
Consequently, according to Eq. (2),
$\rho_B^2=1+\sum\limits_{i=1}^2(m_i/\Lambda_i)^2$.
With the use of the given above values for model parameters we obtain
$\rho_M^2=1,~\rho_B^2=1.5$. The results of the calculations of heavy
meson and heavy baryon observables are
given in tables 1-4. We see that the results of our two-parameter fit
agree well with the existing experimental data and the predictions of
other models.
\vspace*{.5cm}
\def\arraystretch{1.2}

\noindent
\hspace*{.4cm}{\bf Table 1.} $f_B$ and $f_D$ decay constants in MeV.

\begin{center}
\begin{tabular}{|c|c|c|c|c|c|c|c|}
\hline
 & Our & UKQCD & Ivanov[12] & Faustov[13]& Shuryak[14]
 & Narison[15]\\
\hline
 $f_D$ & 226 & $200\pm 30$ & 118$\pm$ 59& 200 &220 &173$\pm$16\\

 $f_B$ & 134 & $180\pm 40$ & 92$\pm$ 40 &120 &140 &182$\pm$18\\
\hline
\end{tabular}
\end{center}
\vspace*{.5cm}

\noindent
\hspace*{1cm}{\bf Table 2.} Branching ratios (in \%) and asymmetry parameters \\
\hspace*{2.8cm} in the decay ${B\rightarrow D(D^*)e\bar\nu}$

\begin{center}
\begin{tabular}{|l|l|l|}
\hline
       & Theory & Experiment \\
\hline
$Br(B\rightarrow D)$ & 2.05 $|V_{bc}/0.04|^2$ & $1.6\pm 0.7$, $1.9\pm 0.5$ \\
\hline
$Br(B\rightarrow D^*)$ & 5.35 $|V_{bc}/0.04|^2$ & $5.3\pm 0.8$, $4.56\pm 0.27$ \\
\hline
$Br(B\rightarrow D^*)/Br(B\rightarrow D)$ & 2.61 & $2.6^{+1.1+1.0}_{-0.6-0.8}$ \\
\hline
${  \alpha_{pol}}$ &1.71 & ${  1.1\pm 0.4\pm 0.2}$  \\
\hline
${  \alpha'}$&0.63 & \\
\hline
${  A_{FB}}$ &0.083&${  0.20\pm 0.08\pm 0.06}$ \\
\hline
${  A_{FB}^T}$&0.20 & \\
\hline
\end{tabular}
\end{center}
\vspace*{.5cm}

\vspace*{.5cm}

\noindent
\hspace*{3cm}{\bf Table 3.} Decay rates of bottom baryons\\
\hspace*{4.8cm}(in $10^{10}$ sec$^{-1}$) for $|V_{bc}|=0.04$

\begin{center}
\begin{tabular}{|l|c|c|c|c|c|}
\hline
Process & [16] & [17] & [2] & [18] & Our \\
\hline
$\Lambda_b^0\rightarrow\Lambda_c^+$   &5.9 &5.1 &5.14 &5.39 &6.02 \\
$\Xi_b^0\rightarrow\Xi_c^+$           &7.2 &5.3 &5.21 &5.27 &6.40 \\
$\Sigma_b^+\rightarrow\Sigma_c^{++}$  &4.3 &    &     &2.23 &2.47 \\
$\Omega_b^-\rightarrow\Omega_c^0$     &5.4 &2.3 &1.52 &1.87 &2.62 \\
$\Sigma_b^+\rightarrow\Sigma_c^{*++}$ &    &    &     &4.56 &5.13 \\
$\Omega_b^-\rightarrow\Omega_c^{*0}$  &    &    &3.41 &4.01 &5.47 \\
\hline
\end{tabular}
\end{center}
\vspace*{.5cm}

\noindent
\hspace*{2.8cm}{\bf Table 4.} Asymmetry parameters for $\Lambda_b$ decay

\begin{center}
\begin{tabular}{|l|c|c|c|c|c|c|}
\hline
~ & $\alpha$ & $\alpha'$ & $\alpha''$ & $\gamma$ & $\alpha_P$ & $\gamma_P$ \\
\hline
Our               &-0.77 &-0.12 &-0.54 &0.55 &0.40 &-0.16 \\
{[18]}            &-0.76 &-0.12 &-0.53 &0.56 &0.39 &-0.17 \\
{[19]}            &-0.74 &-0.12 &-0.46 &0.61 &0.33 &-0.19 \\
\hline
\end{tabular}
\end{center}

\newpage

{\Large\bf Acknowledgements}

\vspace*{.3cm}
\noindent
M.A.I, V.E.L and A.G.R thank Mainz University for the hospitality
where a part of this work was completed. A.G.R. thanks
T.Kopaleishvili for the introduction to the baryon BS equation.
This work was supported in part by the Heisenberg-Landau Program,
by the Russian Fund of Basic Research (RFBR) under contract
96-02-17435-a, the State Committee of the Russian Federation for
Education (project N 95-0-6.3-67, Grant Center at S.-Petersburg State
University) and  by the BMBF (Germany) under contract 06MZ566.
J.G.K. acknowledges partial support by the BMBF (Germany) under contract
06MZ566.

\vspace*{0.7cm}
{\Large\bf References}

\vspace*{.3cm}
\baselineskip 18pt
\noindent
1. M. Neubert, Phys. Rep. {\bf 245} (1994) 259.

\noindent
2. J.G. K\"orner, D. Pirjol and M. Kr\"amer, Prog. Part. Nucl. Phys. {\bf 33} (1990) 757.

\noindent
3. T. Kopaleishvili and A. Rusetsky, Nucl. Phys. {\bf A587} (1995) 758;\\
\hspace*{.3cm}   Yad. Fiz. {\bf 59} (1996) 914.

\noindent
4. N. Brambilla, P. Consoli and G.M. Prosperi, Phys. Rev. {\bf D50} (1994) 5878.

\noindent
5. H.-Y. Jin, C.-S. Huang and Y.B. Dai, Z. Phys. {\bf C56} (1992) 707;\\
\hspace*{.3cm}   G. Zoller et al., Z. Phys. {\bf C68} (1995) 103;\\
\hspace*{.3cm}   A. Abd-el Hady et al., Phys. Rev. {\bf D55} (1996) 4629;

\noindent
6. R.N.Faustov, V.O.Galkin, A.Yu.Mishurov, Phys.Rev. {\bf D53} (1996) 1391, 6302.

\noindent
7. X.-Y. Guo, T. Muta, Phys. Rev. {\bf D54} (1996) 4629.

\noindent
8. J.G. K\"{o}rner and M. Kr\"{a}mer, Z. Phys. {\bf C55} (1992) 659;\\
\hspace*{.3cm}   F. Hussain, J.G. K\"{o}rner and G. Thompson, Ann. Phys. {\bf 206} (1991) 334;

\noindent
9. M.A. Ivanov, J.G. K\"{o}rner, V.E. Lyubovitskij and A.G. Rusetsky,\\
\hspace*{.3cm}    Preprints MZ-TH/97-15, MZ-TH/97-21 (1997).

\noindent
10. S. Chakrabarty et al., Prog. Part. Nucl. Phys. {\bf 22} (1989) 43;\\
\hspace*{.5cm}    A.N. Mitra and I. Santhaham, Few-Body Syst., {\bf 12} (1992) 41.

\noindent
11. Yu-bing Dong, Jun-chen Su and Shi-shu Wu, J. Phys. {\bf G20} (1994) 73.

\noindent
12. M.A. Ivanov, O.E. Khomutenko,  Yad. Fiz. {\bf 53} (1991) 539.

\noindent
13. V.A. Galkin, A.Yu Mishurov and R.N. Faustov, Yad. Fiz. {\bf 53} (1991) 1676.

\noindent
14. E. Shuryak, Nucl. Phys. {\bf B 198} (1982) 83.

\noindent
15. S. Narison, Phys. Lett. {\bf B198} (1987) 104.

\noindent
16. R. Singleton, Jr., Phys. Rev. {\bf D43} (1991) 2939.

\noindent
17. H.-Y. Cheng and B. Tseng, Phys. Rev. {\bf D48} (1993) 4188.

\noindent
18. M.A. Ivanov, V.E. Lyubovitskij, J.G. K\"{o}rner and P. Kroll,\\
\hspace*{.5cm}    Phys. Rev. {\bf D56} (1997) 348.

\noindent
19. B. K\"onig et al., Preprint DESY 93-011 (1993)

\end{document}